\documentclass[usenatbib]{mn2e}

\input psfig.sty

\title[Mysterious Absence of Neutral Hydrogen]{The Mysterious Absence of 
Neutral Hydrogen within One Mpc of a Luminous Quasar 
at Redshift 2.168}
\author[P. J. Francis \& J. Bland-Hawthorn]{
Paul J. Francis$^{1,2}$ and  
Joss Bland-Hawthorn$^{3}$\thanks{E-mail:
pfrancis@mso.anu.edu.au (PJF); jbh@aaoepp.aao.gov.au (JBH)} \\
$^{1}$Research School of Astronomy and Astrophysics, the Australian
National University, Canberra 0200, Australia\\
$^{2}$Joint Appointment with the Department of Physics,
Faculty of Science, the Australian National University\\
$^{3}$Anglo-Australian Observatory, P.O. Box 296, Epping, NSW 2121, Australia
}
\begin{document}

\maketitle

\label{firstpage}

\begin{abstract}

The intense UV radiation from a highly luminous QSO should excite fluorescent
Ly$\alpha$ emission from any nearby neutral hydrogen clouds. We present a
very deep narrow-band search for such emission near the z=2.168
quasar PKS 0424-131, obtained with the Taurus Tunable Filter on the
Anglo-Australian Telescope. By working in the UV, at high spectral resolution
and by using charge shuffling, we have been able to reach surface brightness
limits as faint as $4.7 \times 10 ^{-19}{\rm erg\ cm}^{-2}{\rm s}^{-1}{\rm
arcsec}^{-2}$.
No fluorescent Ly$\alpha$ emission is seen, whereas QSO absorption-line
statistics suggest that we should have seen $\ga 6$ clouds, unless the clouds are
larger than $\sim 100$ kpc in size. Furthermore,
we do not even see the normal population of Ly$\alpha$ emitting galaxies 
found by other surveys at this redshift. This is very different from
observations of high redshift radio galaxies, which seem to be surrounded by
clusters of Ly$\alpha$ emitters. We tentatively conclude that 
there is a deficit of neutral hydrogen close to this quasar, perhaps 
due to the photo-evaporation of nearby dwarf galaxies.

\end{abstract}

\begin{keywords}
Diffuse radiation --- intergalactic medium --- quasars: absorption-lines 
--- quasars: individual: PKS 0424-131 --- galaxies: high-redshift
\end{keywords}

\section{Introduction}

QSO absorption-line statistics tell us that the high redshift ($z>2$) universe
contains many clouds of neutral hydrogen: clouds containing as many baryons as
all the stars today \citep[eg.][]{pei99,per03}. The
absorption-line statistics do not, however, tell us the sizes, shapes 
and physical nature of these clouds. 

There are two broad schools of thought on the nature of these gas clouds. They
could be large ($> 10$ kpc) disk-like structures, associated with young disk
galaxies \citep[eg.][]{pro98}. Alternatively they could lie in tiny 
irregular proto-galactic fragments \citep[eg.][]{hae98}. At intermediate 
redshifts, Mg~II absorbers
(which are probably the same thing as Lyman-limit systems) seem to lie in
large circular structures surrounding disk galaxies 
\citep[eg.][]{ber91,che01,ste02}, and extending out many tens of Kpc. 
But at higher redshift, associations between galaxies and neutral hydrogen
remain obscure \citep[eg.][]{mol02,ade03,fra04}.

Perhaps the most direct way to constrain the nature of these clouds would 
be to measure their typical sizes. One way to do this is to use close pairs 
of QSOs and look for the fraction of absorption-lines seen in both spectra 
\citep[eg.][]{fol84,fra93,sme95,cro98,rau99,imp02}. These observations show 
that the lower column density Ly$\alpha$ forest lines, and perhaps high 
ionisation metal-line systems have sizes of at least
tens of kpc. There are, however, too few such QSO pairs to put useful constraints on
damped Ly$\alpha$ and Lyman-limit absorbers, though there is some evidence that the
probably related low-ionisation metal-line systems are not so large.

Is it possible to directly image these clouds? Their 21cm emission is 
too faint for current telescopes.  Fluorescent Ly$\alpha$ emission is 
a more promising technique. Originally suggested 
by \citet{hog87}, the idea is that the ultraviolet (UV) background
at high redshifts will ionise any neutral gas clouds, and cause them to
emit the Ly$\alpha$ line. \citet{gou96} showed 
that the number of Ly$\alpha$ photons emitted will be $\sim 60$\%
of the number of incident ionising photons.

One could thus use either narrow-band imaging or spectroscopy to search for this
fluorescent emission. Unfortunately, the predicted surface brightnesses
($\sim 3 \times 10^{-20} {\rm erg\ cm}^{-2}{\rm s}^{-1}{\rm arcsec}^{-2}$ at
redshift 4) are very faint, and have not yet been reached 
\citep[eg.][]{bun98,fra01}.

In this paper, we use three tricks to bring this faint emission within reach:

\begin{enumerate}

\item Work in the UV. Most of the deepest narrow-band Ly$\alpha$ searches
made to date have been done at redshifts 3 \ -- \ 7, where the Ly$\alpha$ line
lies at at visible or red optical wavelengths 
\citep[eg.][]{bun98,ste00,ouc03,hu04}. The observed surface brightness of a
given object is, however, a very strong function of redshift. Move a given
Ly$\alpha$ emitting cloud from redshift two out to redshift five and the
observed flux per square arcsecond recorded at the Earth drops by more than an
order of magnitude. This effect more than compensates for the increased
sensitivity of most CCDs at red wavelengths. So if the aim is to detect low
luminosity sources, you are better off working at the lowest redshift possible.
The very faint and stable sky background also helps. In addition, working in the UV minimises the number of
strong emission lines in foreground sources that could impersonate Ly$\alpha$.
Luckily, the Anglo-Australian Telescope has recently commissioned an EEV CCD 
which achieves close to the theoretical maximum in blue 
quantum efficiency ($\approx 90$\% at B; see http://www.aao.gov.au/cgi-bin/ttf).

\item High spectral resolution. Most narrow-band Ly$\alpha$ imaging to date
has used custom monolithic interference filters, operated in converging beams
\citep[eg.][]{cam99,ste00,pal04}. This set-up has many advantages, allowing use
of a wider range of telescopes and large fields of view at prime focus.
Unfortunately, it forces the use of a relatively wide spectral bandpass, typically
50 --- 150\AA . We use a much narrower (7 \AA ) bandpass, reducing the sky
background substantially, albeit at substantial cost to the co-moving volume
surveyed.

\item Work near a QSO. If we image a region close to a luminous QSO, its UV
emission will increase the ionisation and hence the Ly$\alpha$ emission of any
nearby neutral hydrogen clouds. A 17th magnitude QSO, for example, will
increase the ionising flux incident on a cloud by a factor of $\sim 5$ at a
distance of one proper Mpc. The price paid for this is that the regions close to
luminous QSOs are not typical parts of the high redshift universe: clustering might 
enhance the number of gas clouds, but as discussed in \S~\ref{dis}, the QSO
may destroy nearby gas clouds. This
technique was suggested by \citet{hai01} for studying gas in the halos of very
high redshift QSOs, and was successfully used by \citet{mol93} to image a damped
Ly$\alpha$ system.

\end{enumerate}

In this paper, we present high spectral resolution narrow-band imaging in 
the near-UV of the volume around the very luminous quasar PKS 0424-131, at 
redshift 2.159. We reach sensitivity limits that should have allowed us to see
the fluorescent Ly$\alpha$ emission from any neutral hydrogen clouds within 
$\sim 1$ Mpc of the quasar. None were seen, and we discuss the reasons for this 
non-detection.

We assume $H_0 = 70 {\rm km\ s}^{-1}{\rm Mpc}^{-1}$, 
$\Omega_{\rm matter} = 0.3$ and $\Omega_{\Lambda} = 0.7$ throughout 
this paper. We quote proper (rather than co-moving) distances throughout.

\section{Observations and Reduction\label{obsred}}

\subsection{Target Selection\label{target}}

We required a target QSO which was as bright as possible, and lay 
at a redshift that placed its Ly$\alpha$ emission within one of the
Anglo-Australia Observatory's intermediate band blocking filters.
More challengingly, the QSO systemic redshift needed to be known to an 
accuracy of better than $10^{-3}$, so that 
we could confidently target Ly$\alpha$ emission from nearby gas clouds
even with a very narrow-band filter.

The best way to measure a QSO systemic redshift with enough precision is
to observe its narrow forbidden emission lines, which are shifted into the
near-IR at these redshifts. As these lines are thought to be emitted at
distances of hundreds of parsecs or more from the QSO nucleus, they should
reflect the redshift of the host galaxy with adequate precision, unlike the
broad emission-lines, with the possible exception of Mg~II (2798\AA ) 
\citep[eg.][]{esp89}.

\citet{esp89} and \citet{mci99} have measured accurate forbidden-line
redshifts, in the near-IR, for small samples of luminous QSOs. The best 
placed of these QSOs for our scheduled observing dates was PKS 0424-131. 
PKS 0424-131 is an optically bright (B=17.6) QSO with a strong narrow 
[O~III] (5007\AA ) emission line seen in the spectrum of \citet{mci99}.
Using this line, a redshift of 2.168 was measured. A consistent redshift was
also measured from its Mg~II (2798\AA ) line. 

PKS 0424-131 is a powerful radio source \citep[1.0 Jy at 1.4 GHz,][]{con98}, with a
radio spectral index ($F_{\nu} \propto \nu^{-0.55}$) that places it close to the
conventional boundary between flat- and steep-spectrum sources. It is a compact
($<0.5$\arcsec ) radio source at 6cm wavelength \citep{bar88}. This QSO has been
the subject of over 100 papers, though little of the data in these papers is relevant to
this project. An archival Hubble Space Telescope (HST) Faint Object Spectrograph spectrum 
shows that there is no Lyman-limit absorption along our line of sight within the region 
we probe. There are Ly$\alpha$ forest lines within this region, and curiously,
an associated metal-line absorption system at
redshift 2.17288 \citep{pet94}: ie. infalling at over $400{\rm km\ s}^{-1}$,
compared to the [O~III] or Mg~II redshifts. There are a variety of faint near-IR sources
seen close to the QSO, either published in \citet{ara94} or seen in the deep archival HST
NICMOS imaging, but no information is available on the redshifts of these sources.

The [O~III] redshift would place Ly$\alpha$ emission at 3851\AA .
Mg~II would place it slightly bluer at 3848\AA , while Ly$\alpha$ emission
at the redshift of the associated absorber would lie at 3857\AA . Our three
bandpasses (\S~\ref{tech})  cover all these possibilities.

We can estimate the black hole mass in this quasar, using the measured width of
the H$\beta$ line \citep{mci99} and the rest-frame 5100\AA\ (observed frame 
K-band) flux from the Two Micron All Sky Survey, using the relations in 
\citet{kas00}. The inferred mass is $6.6 \times 10^9 {\rm M}_{\sun}$. 
Extrapolating still 
further, using the results of \citet{fer02}, we infer a dark halo mass of 
$5 \times 10^{14} {\rm M}_{\sun}$. Note that both numbers are based on 
large extrapolations from the existing low redshift data,
and should therefore be regarded with caution.

The Galactic extinction along this sight-line, from \citet{sch98}, is
$A_B = 0.3$ mag.

\subsection{Technical Details\label{tech}}

Our observations were carried out on the nights of 2003 December 22 -- 25,
using the Taurus Tunable Filter \citep[TTF, ][]{bla98} on the Anglo-Australian 
Telescope (AAT). The TTF is an etalon used in low order, with a greatly 
extended physical scan range, to give a 
relatively wide and tunable bandpass. Conditions were photometric 
throughout, and the median seeing was 1.8\arcsec . A blue sensitive EEV CCD was
used as the detector, the pixel size being 0.33\arcsec\ on the sky.

We observed a field 500\arcsec\ East-West by 235\arcsec\ North-South, centered
on the QSO PKS 0424-131. The filter bandpass was a Lorentzian of full width at
half maximum transmission 7\AA . A narrower bandpass would have improved the
sensitivity, but would have required a narrower bandpass blocking filter than
was available to block other orders. The field was observed with the TTF set to
three different central wavelengths: 3851\AA\ (Ly$\alpha$ at the QSO redshift),
and wavelengths 6\AA\ blueward and redward of this. This is the wavelength at
the centre of the field; it shifts to the blue quadratically towards the edge
of the field. The shift at a radius of 2.5\arcmin\ is 6\AA\ to the blue.
Other interference orders were blocked by an 170\AA -band blocking filter,
centered at 3900\AA .

Wavelength calibration proved a challenge at these UV wavelengths. Extensive
experimentation showed that the best calibration lamp was a low pressure Xe
lamp, which gave three measurable lines within the blocking filter bandpass.
Wavelength calibration scans were performed every three hours during the night.
Wavelength shifts between scans were typically 1\AA\ or less.

We used the charge-shuffling technique linked to frequency switching 
\citep[][]{bla98, mal01} to minimise
systematic errors in sky subtraction. We only observed with a quarter of the CCD
at a time. Every minute, the charge was shuffled up and down the chip,
synchronised with shifts in the TTF plate spacing. The result is three images
on the same chip, one at each of our three central wavelengths, observed at
essentially the same time and through the same pixels. We ran for 45 minutes
between reading out the CCD, giving a 15 minute exposure at each of the three
central wavelengths. The pointing of the telescope was shifted by $\sim
10$\arcsec\ after every read-out, to further improve sky subtraction.

The total exposure time at each central wavelength was 24,420 sec. Flux
calibration was obtained using the same instrumental set-up, and observing 
three spectrophotometric standard stars. Twilight flat field frames were
obtained with the etalon removed from the beam, but with the blocking filter in
place.

\subsection{Data Reduction}

The reduction and analysis of tunable filter images is discussed in detail
in \citet[][]{jon02}.
The raw data showed many horizontal bars of elevated charge, probably due to
charge trapping sites in the CCD. To remove these, a master night sky flat was
produced by combining all the data frames unshifted, and using rejection of high
pixels to remove stars. This worked very well and reduced the bars to
undetectable levels.

The data frames were flat fielded and bias subtracted, and then aligned, using
all bright stars in the field to compute offsets. Integer pixel shifting was
used, to avoid introducing spurious correlations between adjacent pixels. The 
shifted
frames were then combined. Pixels discrepant at the 5$\sigma$ level, together
with their neighbors, were removed from the final coadded image: this did an
excellent job of removing cosmic ray hits. Observing conditions were so uniform
that no weighting of the coadded frames was required.

The three different images with their different wavelengths were then
extracted from the coadded CCD frame. The average of two images was used as the
``off-band'' for the other image image, and subtracted from it to provide a 
difference image.

The noise properties of the data show that the sky subtraction technique had
succeeded in minimising systematic errors. The standard deviation of the sky
pixel values in the coadded image was within 10\% of that predicted by
dividing the standard deviation in an individual image by the square root of the
number of frames combined. Noise properties were also highly uniform across the
field. When the coadded image was binned up, once again the
standard deviation went down as $\sqrt{n}$, where $n$ refers to the number
of detected photoelectrons, even up to very large binning
factors ($30 \times 30$ pixels).

\section{Results}

\begin{figure*}
\psfig{figure=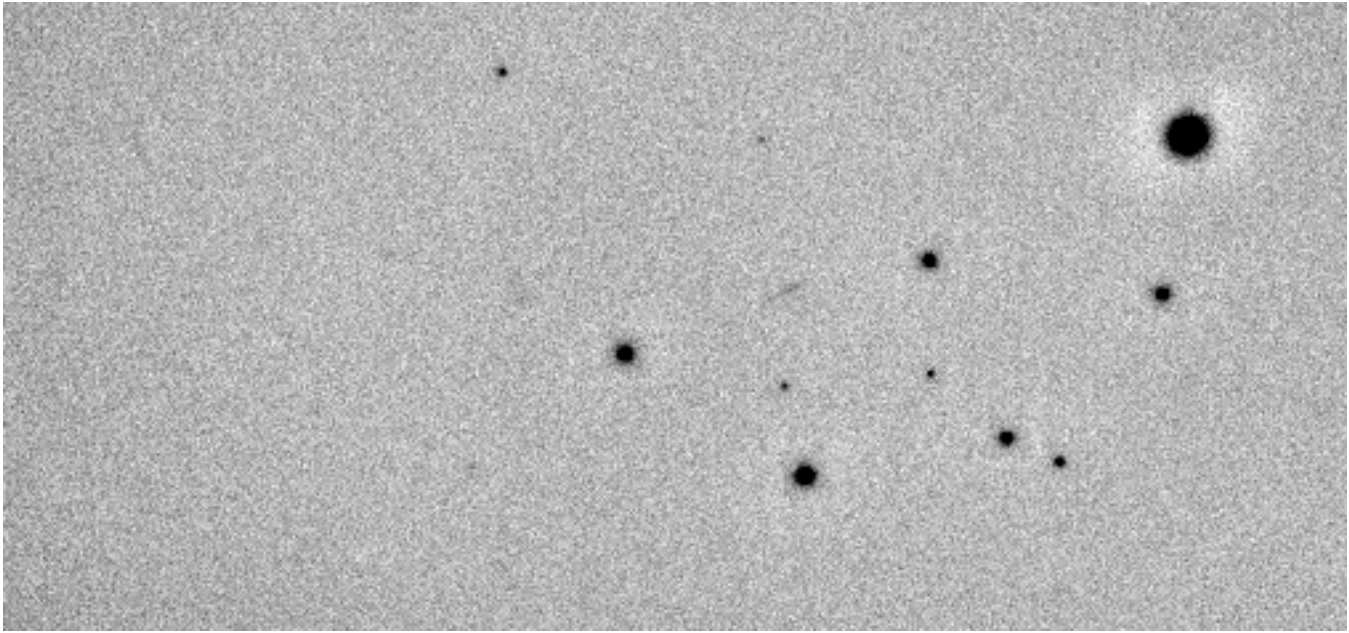}
  \caption{
Image of our field centred at 3845\AA . North is upwards and East to
the left. The field is 500\arcsec\ east-west and 235\arcsec\ 
north-south. PKS 0424-131 is the bright source slightly below the centre of the
field.\label{im1}
  }
\end{figure*}

\begin{figure*}
\psfig{figure=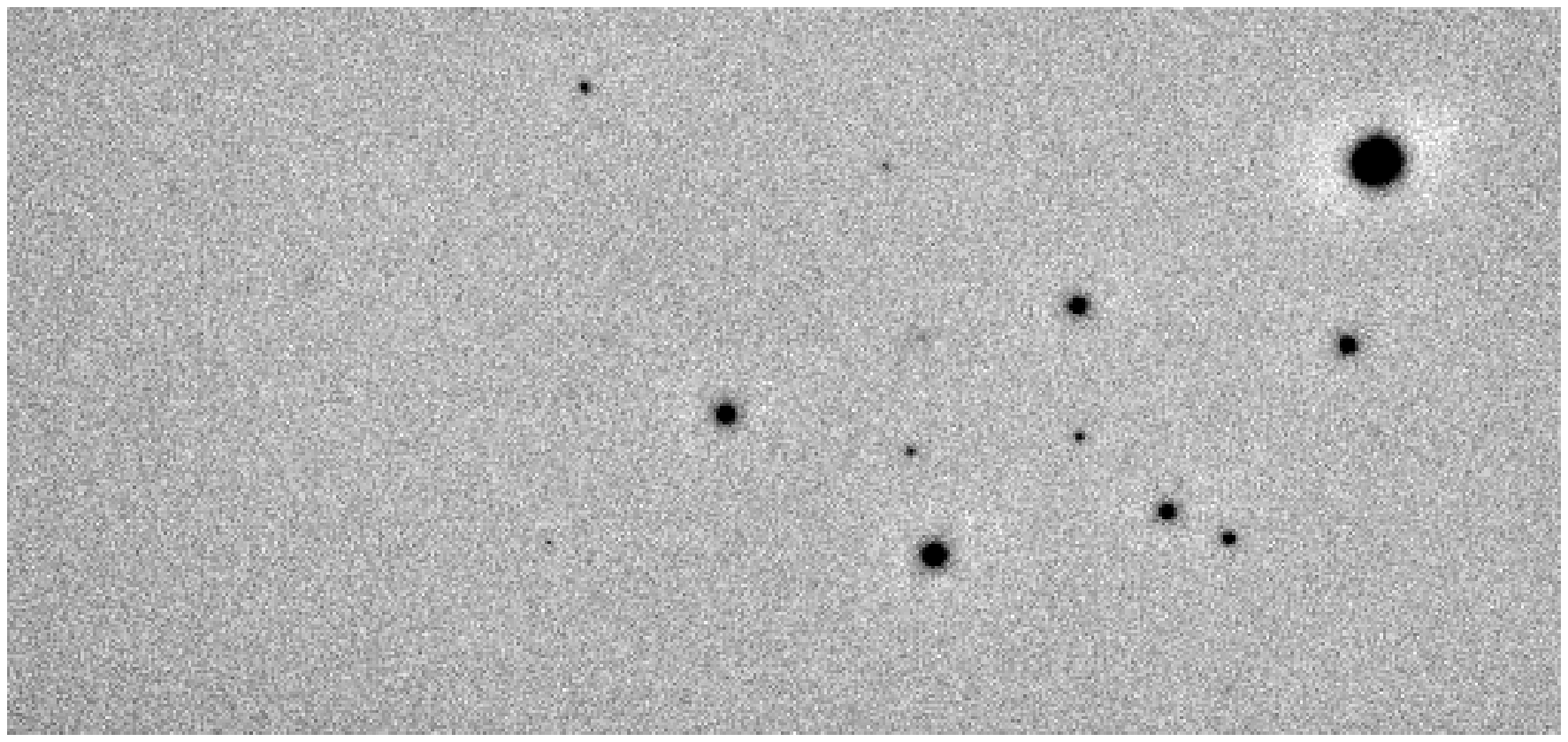}
  \caption{
Image of our field centred at 3851\AA . Details as in Fig~\ref{im1}.
\label{im2}
  }
\end{figure*}

\begin{figure*}
\psfig{figure=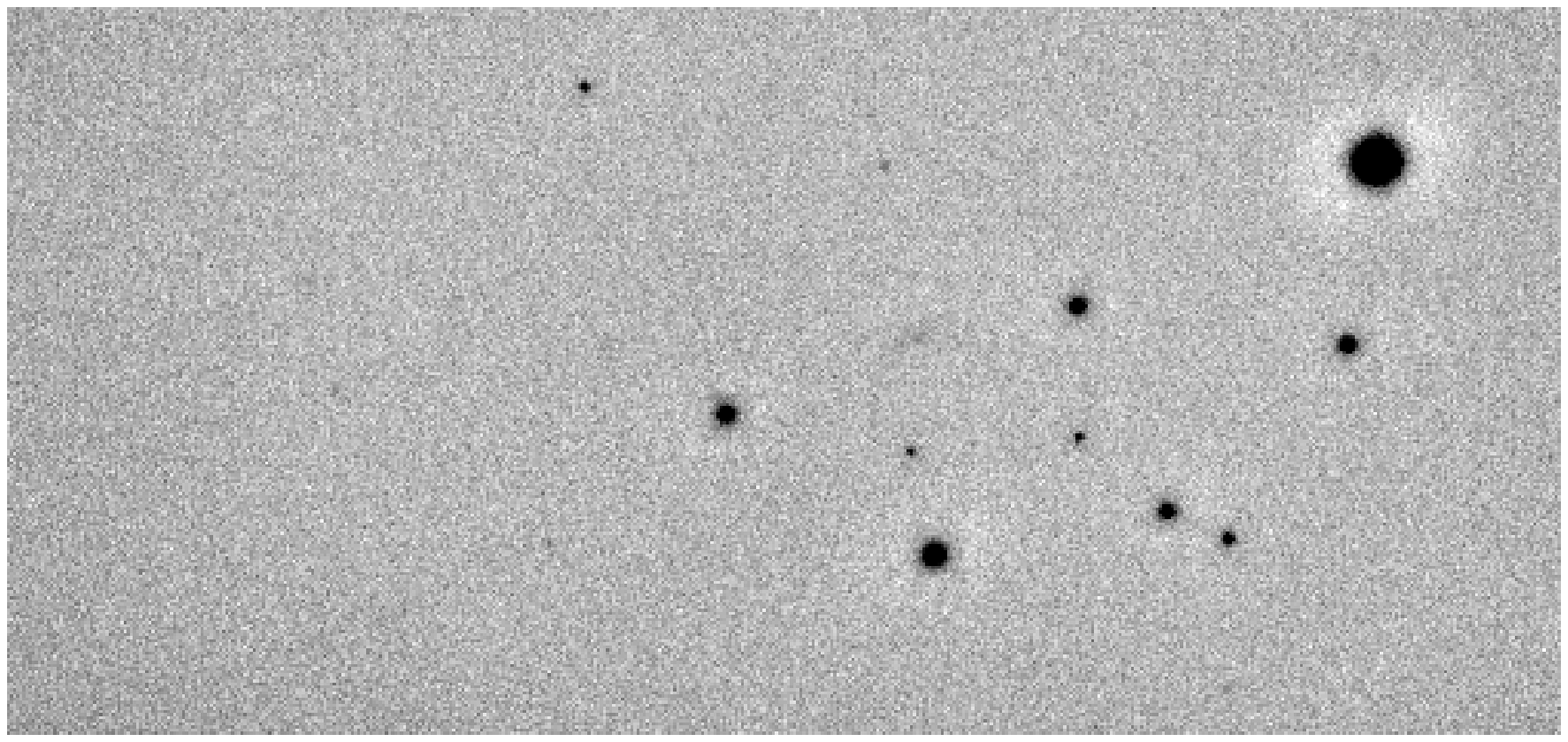}
  \caption{
Image of our field centred at 3857\AA . Details as in Fig~\ref{im1}.
\label{im3}
  }
\end{figure*}

The three narrow-band images are shown in Figs~\ref{im1}, \ref{im2} and
\ref{im3}. The images were
searched for narrow-band emission by blinking them, by visual inspection 
of the difference images, and using the Source Extractor (SExtractor)  program 
\citep{ber96}. The excellent noise properties of the images allowed us to
search for low surface brightness features both by binning up the data and by
looking for large numbers of contiguous pixels as little as 0.2 standard deviations 
above the background level. 

No sources were found with excess flux in any of the narrow-band filters
significant at the $5 \sigma$ level, on any scale. Our $5 \sigma$  sensitivity 
limit for unresolved ($< 2$\arcsec ) sources corresponds to a line flux of
$9.6 \times 10^{-18}{\rm erg\ cm}^{-2}{\rm s}^{-1}$. For sources larger than
$3 \times 3$\arcsec , our $5 \sigma$ surface brightness limit is
$1.6 \times 10^{-18}{\rm erg\ cm}^{-2}{\rm s}^{-1}{\rm arcsec}^{-2}$. This
drops to $4.7 \times 10^{-19}{\rm erg\ cm}^{-2}{\rm s}^{-1}{\rm arcsec}^{-2}$
for sources larger than $10 \times 10$\arcsec . Galactic absorption at this
wavelength is $\sim 0.33$ mag \citep{sch98}.

\section{Predicted Source Properties}

The Ly$\alpha$ sources could be ionised by the quasar (externally ionised) or
by UV sources within them (internally ionised). We consider these in turn.

\subsection{Externally Ionised Hydrogen\label{pgas}}

\subsubsection{Surface Brightness\label{flux}}

How bright should the fluorescent Ly$\alpha$ emission from a neutral 
hydrogen cloud lying close to a bright QSO be?
Let us approximate the spectrum of a QSO as a power-law, of the form
$L_{\nu} = L_0 (\nu / \nu_0 )^{\alpha}$
where $L_{\nu}$ is the luminosity per unit frequency, and $L_0$ is this 
luminosity at a
particular frequency $\nu_0$. For PKS 0424-131, we measured a  flux of
$4.3 \times 10^{-16}{\rm erg\ cm}^{-2}{\rm s}^{-1}$\AA$^{-1}$ at a
wavelength of 5000\AA , which corresponds at this redshift to a luminosity of
$5.5 \times 10^{23}{\rm W\ Hz}^{-1}$ at rest-frame $6 \times 10^{14}{\rm Hz}$
(correcting for Galactic extinction). This is very consistent with the various
brightness measurements of this quasar in the literature: we see no evidence 
for variability. In the rest-frame far-UV, QSOs have $\alpha \sim -0.7$
\citep{fra91}. The number of ionizing photons emitted per unit time is thus:

\begin{equation}
N = L_0 \int_{\nu_{LL}}^{\infty} \frac{1}{h \nu} \left(\frac{\nu}{\nu_0} 
\right)^{\alpha} d\nu 
= \frac{-L_0}{\alpha h} \left( \frac{\nu_{LL}}{\nu_0} \right)^{\alpha}
\end{equation}

where $\nu_{LL}$ is the frequency of the Lyman limit ($3.29 \times 10^{15}$ Hz).

Let us assume that this photon flux impacts face-on upon a plane-parallel cloud of neutral
hydrogen, lying at a distance $r$ from the QSO. If the cloud is optically thick
(typically $N_H > 10^{17.2}{\rm cm}^{-2}$), all ionising photons will be
absorbed. A fraction $\varepsilon$ will then be re-radiated as Ly$\alpha$
photons. \citet{gou96} showed that $\varepsilon \sim 0.6$. The Ly$\alpha$
luminosity emitted by a neutral hydrogen cloud of area $A$ is thus:

\begin{equation}
L_{Ly \alpha} = \frac{\varepsilon}{4 \pi r^2} \frac{L_0}{\alpha}
\nu_{Ly\alpha} \left( \frac{\nu_{LL}}{\nu_0} \right)^{\alpha}
\end{equation}

Given the observed brightness of PKS 0424-131, assuming $\alpha = -0.7$,
$\varepsilon = 0.6$, and that the cloud surface is orthogonal both to the quasar
light and to our sight-line, the predicted surface brightness is thus:

\begin{equation}
1.2 \times 10^{-18} \left(\frac{r}{ 1~{\rm Mpc}} \right)^{-2} {\rm erg\ cm}^{-2}{\rm
s}^{-1}{\rm arcsec}^{-2}
\end{equation}

Gas clouds whose surfaces are at an oblique angle to the incident photon flux
will appear fainter, while those at an oblique angle to our line of sight will
appear brighter: the two effects should cancel on average, but introduce a
dispersion into the predicted surface brightnesses. We are ignoring
radiative transfer within the cloud. The Ly$\alpha$ transition can have an
enormous optical depth, but most of the clouds we will be seeing will be only
marginally optically thick in the Lyman limit, so this effect may not be
too severe. It will be severe for any damped Ly$\alpha$ systems lying near
the QSO: we may only be able to see these if they lie on the far side of the
QSO, so we are directly observing the illuminated face.

Given these predicted surface brightnesses, and our measured sensitivity limits,
we should be able to detect neutral hydrogen clouds that subtend $1 \times 
1$\arcsec\ (8.2 proper kpc) out to $r \sim 300$ kpc from the quasar. Larger 
clouds should be detectable at larger radii: clouds subtending $3 \times
3$\arcsec\ should be detectable out to $r \sim 740$ kpc while clouds subtending
$10 \times 10$\arcsec\ should be detectable out to $r \sim 1350$ kpc.

\subsubsection{Source Density\label{dens}}

We are sampling a much smaller co-moving volume than most previous narrow-band
Ly$\alpha$ surveys: is there a risk that we might not find anything in so small
a volume?

QSO absorption-line statistics tell us the typical number of Lyman limit systems
found per unit redshift along a QSO sight-line. \citet{per03} have compiled
data from several surveys and give a number of Lyman-limit systems per unit
redshift at z=2.16 of $N(z) = 1.6$. In this section, we assume that the region close to the
QSO is typical of the universe at this redshift (but see \S~\ref{dis}).

Let us now assume that Lyman-limit systems are all face-on squares of
side-length $r$.
Consider a square region of the sky, of size $D \times D$, observed with a
narrow-band filter sensitive to Ly$\alpha$ emission over a redshift range
$\Delta z$. The probability of any given sight-line through this volume
intersecting a Lyman-limit system is $P = N(z) \Delta z$. This should be equal
to the fraction of the cross-sectional area of this region covered by the
Lyman-limit systems; ie. $n r^2/D^2$, where $n$ is the number of Lyman-limit
systems in the region, and $P \ll 1$. Thus:

\begin{equation} 
n = N(z) \Delta z \left( \frac{D}{r} \right)^2
\end{equation}

\begin{figure*}
 \psfig{figure=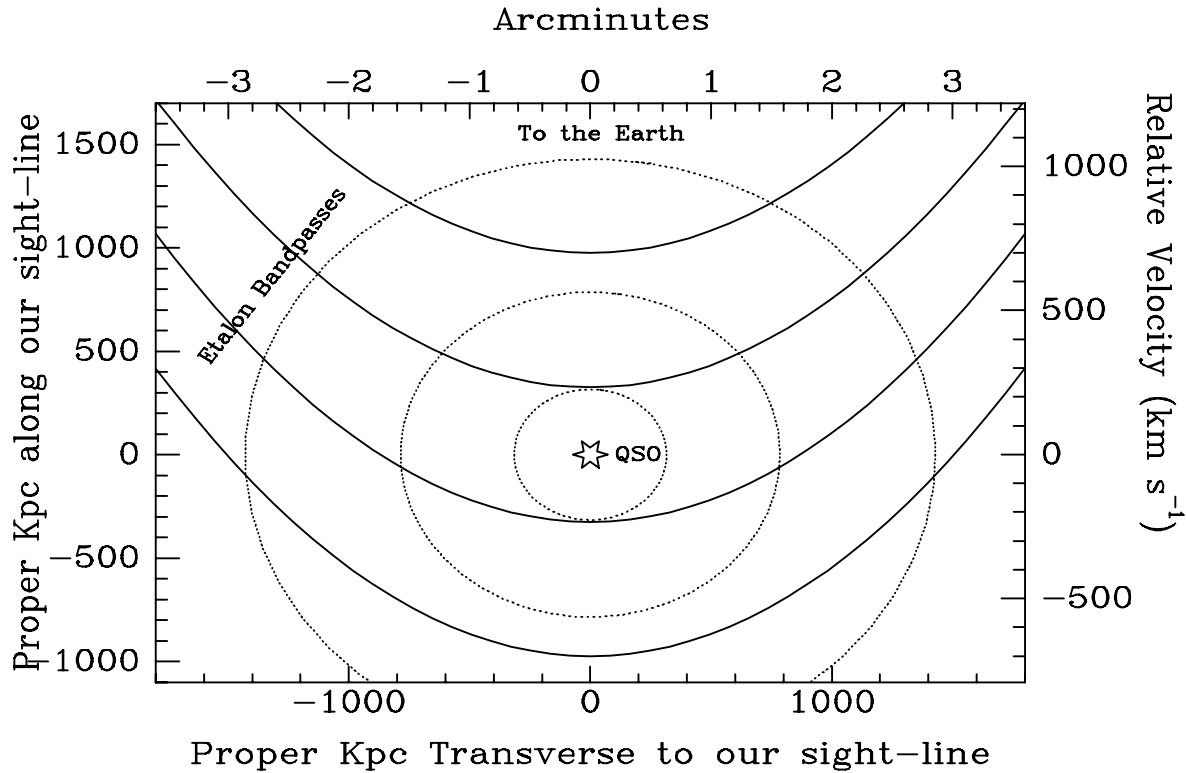}
  \caption{
  The 3D geometry of the volume in which we are sensitive to neutral hydrogen
clouds. The solid lines show the volumes sampled by our three different
central wavelength images. The dotted circles show the volumes out to which we
should be able to find clouds subtending $1 \times 1\arcsec$ (smallest circle),
$3 \times 3\arcsec$ and $10 \times 10\arcsec$ (largest circle). 
1000 Kpc equals two transverse arcminutes at this redshift. A cloud
subtending $1 \times 1\arcsec$ would be unresolved at our $1.8\arcsec$ 
seeing, which has been taken into account in computing the radius out to 
which it would be detectable.\label{geom}
  }
\end{figure*}

The 3D geometry of the region we survey (ignoring peculiar motions) is 
shown in Fig~\ref{geom}. Between
our three central wavelength images, we fully sample a sphere around the
QSO of radius 740 kpc (the radius to which we are sensitive to clouds 
subtending $3 \times 3\arcsec$ or larger), and sample most of a sphere of 
radius 1430 kpc (in which we are sensitive to clouds subtending $10 \times
10\arcsec$ or larger; \S~\ref{flux}).

Could we miss galaxies because of their peculiar motions in the potential
well of a possible cluster surrounding PKS 0424-131? Between the three
images we cover a velocity range
of $\sim 1500 {\rm km\ s}^{-1}$, so this should not be a problem unless
there is a very massive cluster indeed. The existence of a cluster this massive
at this redshift would be interesting in itself \citep[eg.][]{ros02}. The
$z=2.16$ cluster identified by \citet{pen00} had two
sub-components, with velocity dispersions of $300$ and $500 {\rm km\ s}^{-1}$,
so either would easily fit within our passband.

Given this geometry, and our field of view (ample for all except the largest
sphere), and making the same assumptions about neutral hydrogen cloud size as in
\S~\ref{flux}, we should have seen $\sim 20$ $1 \times 1\arcsec$ clouds, 
$\sim 25$ $3 \times 3\arcsec$ clouds, or around 6 $10 \times 10\arcsec$ clouds. Only if the
clouds were larger than 20 arcsec (160 kpc) would the expected number fall below one.

\subsection{Internally Ionised Hydrogen\label{pgal}}

Many authors have shown that at sufficiently faint flux levels, a substantial
population of Ly$\alpha$ emitting sources is seen at high redshift
\citep[eg.][]{cow98,pas98,kud00}. As these sources are seen far from any QSOs,
the ionisation must be generated locally, probably by weak AGNs or star 
formation within the galaxies.

We estimated how many of these sources we should see by comparing to the
deepest existing Ly$\alpha$ surveys. The best match to our survey, both in
redshift and depth, is that of \citet{pas98}. Their flux limit is very
comparable to ours, and the co-moving volume they survey more than twice ours.
Allowing for this difference, their results predict that we should see $\sim 6$
Ly$\alpha$ emitting sources. \citet{kud00} surveyed a much larger volume
at z=3.1, to a luminosity limit about four times brighter than ours. If we
assume that the surface density of Ly$\alpha$ emitters increases by a factor of
five for each magnitude we go deeper \citep{pal04}, their results also
predict that we should see $\sim 5$ sources. Similar numbers can be obtained by
extrapolating from the results of \citet{cow98}.

In practice, these numbers should be lower limits, as the region close to
a luminous QSO is not a random part of the early universe. Work on high redshift
radio galaxies has shown that they are surrounded by enhanced numbers of
Ly$\alpha$ emitting sources \citep[eg.][and references therein]{ven02}.
\citet{kur00} and \citet{pen00} searched for Ly$\alpha$ emitting sources around 
a radio galaxy at an almost identical redshift to our source, and down to a 
flux limit very comparable to our own. They found at least 14 Ly$\alpha$ 
emitting sources and a faint QSO, all within 1.5 projected Mpc of the radio 
galaxy. Comparable clusters of Ly$\alpha$ emitting galaxies are seen around
other high redshift radio galaxies \citep[eg.][]{lef96,pas96,ven02}.

Could the Ly$\alpha$ emitters seen near these radio galaxies simply be neutral
hydrogen clouds photoionised by UV from the AGN, rather than galaxies? While we
see no UV from these radio galaxies, it may be emerging along an ionisation cone
in some other direction. The Ly$\alpha$ sources are not, however, aligned in any
particular area, or close to the radio jet axis, and the concealed nucleus 
would have
to be far brighter in the UV than typical QSOs at this redshift. It therefore
seems likely that these are indeed ionised internally.

We therefore conclude that we should have expected to see $\ga 10$ internally
ionised Ly$\alpha$
emitting sources near our quasar, whereas we actually saw none.

\section{Discussion: Why Didn't We See Anything?}

We showed in \S~\ref{pgas} that if the region surrounding PKS 0424-131 were
typical, we should have expected to have seen the fluorescent Ly$\alpha$
emission from a considerable number of clouds. We should also have seen
internally ionised clouds. Instead, we saw nothing.

Could this just be a statistical fluke? Assuming Poisson statistics, the probability
of seeing no sources given an expected number $n$ is $< 1$\% for $n>4$ and $<0.1$\%
for $n>6$.  Our non-detection is thus reasonably significant, though clearly observations
of several other luminous
QSOs would be needed to test this. There have, however, been very few 
comparable observations. \citet{cam99} imaged a pair of
QSOs at $z=2.58$, and  found 37 candidate Ly$\alpha$ emitters, but they have no
spectral confirmation, their QSOs are much fainter than ours, and they used a
130\AA\ bandpass filter so the candidates may well lie at a somewhat 
different redshift
to the QSOs. \citet{hu96} and \citet{hu96a} found three luminous Ly$\alpha$ 
sources near two bright QSOs, but both QSOs lie at very different redshifts from
ours ($z> 4.5$). The only
observations comparable to ours were those of \citet{ber99}, who observed the
$z=2.238$ QSO in the Hubble Deep Field South down to a flux limit about double
ours. They found three candidate Ly$\alpha$ emitting sources, though their large
(86 \AA ) bandpass means that they too may lie at a different redshift, and they
have not published spectral confirmation of these sources. None of these QSOs have
radio luminosities approaching that of PKS 0424-131.

We thus conclude that there is no compelling evidence in the literature for
or against a deficit of faint Ly$\alpha$ emitters close to luminous, high
redshift QSOs.

In this remainder of this section, we will assume that this deficit of
Ly$\alpha$ emitters is a generic feature of extremely luminous QSOs at $z \sim
2$, and we will ask how such a deficit could be explained.

\subsection{Lack of External Ionisation}

How could we avoid the quasar PKS 0424-131 from ionising any neutral hydrogen
clouds in its vicinity?

If it had only recently switched on its UV emission (at the
time when the continuum radiation we see from it was being emitted), there may
not have been time for this radiation to reach nearby gas clouds, ionise them
and for the Ly$\alpha$ emission to reach the Earth. This would require that PKS
0424-131's current active phase be younger than $\sim 10^6$ years. This is
shorter than most estimates of the total lifespan of luminous QSOs
\citep[eg.][ and refs therein]{mar03}. This total lifespan could, however, be
divided up into several short active phases, or we could have been unlucky
enough to have chosen a quasar very early in its active phase.

As noted in \S~\ref{target} it is possible that PKS 0424-131 is a compact steep
spectrum quasar. If this is the case, the upper limit on the size of the radio
source (0.5\arcsec ), combined with the typical expansion speed of jets
\citep{bic03} allows us to place an upper limit on the age of the jet:
$\sim 10^5$ years. If this radio source is beamed towards us, however, the size
places no limit on the age. We plan Very Long Baseline Interferometry (VLBI)
observations to discriminate between these two possibilities.

The calculations in \S~\ref{dens} assumed that the UV emission from PKS 0424-131
is isotropic. This is known not to be the case for many nearby Seyfert nuclei,
where a ``torus'' intercepts $\sim 50$\% of the ionising flux. If this were the
case here, the expected numbers of externally ionised hydrogen clouds would be
fewer than we predict. Note however that recent X-ray observations
suggest that a smaller fraction of luminous QSO UV emission is blocked by 
tori at high redshifts \citep[eg.][]{ste03}.  Observations of the transverse 
proximity effect in two sQSO \citep{lis01,jak03} provide further evidence that the UV radiation 
from  high redshift escapes into substantial solid angles.
Furthermore, we know from 
nearby Seyferts that up to 5\% of the ionizing radiation field can be
scattered almost isotropically by a warm nuclear wind \citep[][]{mil91,bla91}.
This would render HI clouds visible over a 100~kpc sphere. Note however that, 
especially if the clouds are large, even a small reduction in the radius out to which 
we can see clouds, or even a narrow obscuring torus can drop the predicted number of
clouds below 4, and hence drop the statistical significance of our null detection below
99\%.

\subsection{Absorption of the Ly$\alpha$ Emission}

The Ly$\alpha$ line is the optically thickest transition in astrophysics, and
hence is easily destroyed. If the ionised neutral hydrogen
clouds are optically thick in Ly$\alpha$, their Ly$\alpha$ 
emission would only escape back towards the QSO. This would mean that we 
could only detect Ly$\alpha$ emission from clouds of the far side of the 
QSO. While this
is likely to be true for any systems with $N_H \gg 10^{18}{\rm cm}^{-2}$ 
lying close to PKS 0424-131, we would expect many of the neutral hydrogen
clouds to have lower optical depths, and hence be visible from all
directions.

The high optical depth means that Ly$\alpha$ photons generated within a
cloud will have to random-walk their way out. If any dust is present, this
gives it multiple chances to absorb the Ly$\alpha$ photons. We know, however,
that the Lyman-limit systems seen in absorption against bright QSOs contain very
little dust: if they were dusty, the background QSO would have been too faint
to observe. There may be a population of dusty absorbers, but they would be
in addition to the population known from QSO absorption-line statistics
\citep[eg.][]{fal93}.

We also know that many high-redshift galaxies emit Ly$\alpha$, despite being
ionised from the inside. Any Lyman-limit systems near the QSO would be ionised
from the outside, enhancing the probability of any Ly$\alpha$ photons escaping.

We therefore tentatively conclude that the high optical depth of Ly$\alpha$ is
unlikely to make all the hydrogen clouds in our field invisible. Even if it does
render any externally ionised clouds undetectable, we should see at least the
space density of internally ionised clouds found elsewhere in the high redshift
universe.

\subsection{Destruction of Hydrogen Clouds near the Quasar\label{dis}}

This leaves us with the third possibility: that we see no Ly$\alpha$ emitters
because there are no neutral hydrogen clouds close to the quasar.

If this very luminous QSO lies in a rich cluster environment, the
surrounding gas might be heated to X-ray emitting temperatures, and the
surrounding galaxies might be deficient in neutral hydrogen, by analogy
with the traditional view of modern galaxy clusters \citep[][]{gio85,cay90}.
However, recent studies show that stripping in clusters today is much
less effective than originally suggested. The most comprehensive survey to
date, incorporating 1900 galaxies in 18 clusters \citep[][]{sol01},
finds that more than one-half of spiral galaxies retain at least half
their HI within 900 $h_{70}^{-1}$ kpc. This is the
scale, roughly 40\% of an Abell radius, over which we expect to see
disks with bright HI disks or halos. A third of the
clusters in the Solanes et al. sample show no HI deficiency.

Furthermore, dense galaxy clusters are not thought to be common 
at such high redshifts \citep{ros02}. 
One would also expect high
redshift radio galaxies to inhabit similar environments to our
quasar, and as noted in \S~\ref{pgal}, overdensities of Ly$\alpha$
emitting sources are found around them. Other candidate high-redshift
clusters also seem to have overdensities of Ly$\alpha$ emitters
\citep[eg.][]{ste00,pal04,fra04}.

Could a wind from the quasar have cleared its environment of neutral
hydrogen, as suggested by \citet{sca04}? \citet{ade03} showed evidence
for a similar effect around Lyman-break galaxies, though the reason for this effect
is unclear \citep[eg.][]{cro02,kol03,bru03}. We note however that
the existing evidence points to an overdensity of neutral hydrogen near AGNs,
not a deficit. Furthermore, such a wind could easily increase the brightness 
of any neutral hydrogen clouds by driving shocks into them \citep{fra01b}. 

The major difference between our quasar and the various high redshift radio
galaxies is its enormous UV luminosity (at least along our line of sight).
One possibility is that the quasar environment contains the normal quantity of
hydrogen, but that this UV radiation has almost completely
ionised it, so it is no longer optically thick in the Lyman-limit and is hence
far less efficient at reprocessing the quasar flux into Ly$\alpha$ emission (the same
effect that produces the proximity effect for Ly$\alpha$ forest clouds).
We can compare the incident quasar flux at 912\AA\ to the UV background flux at
this wavelength \citep[$\sim 7 \times 10^{-22} {\rm erg\ cm}^{-2}{\rm Hz}^{-1}{\rm
Sr}^{-1}$,][]{sco02}. Given the observed brightness of the QSO, assuming a
typical QSO spectrum \citep{fra91}, and allowing for the fact that the UV
background will impact both sides of a gas cloud while the quasar illuminates
only one side, we find that at 1 Mpc from the quasar, its flux exceeds the UV
background by a factor of $\sim 3$. At closer distances, where we would be
seeing small clouds, the factor is much larger. Note that the ratio of fluxes at
912\AA\  underestimates the ratio of ionising flux as the UV background is more
filtered by intervening material.

Given this enormous increase in ionising flux, how much might the number of
optically thick externally ionised clouds decrease? We can crudely estimate this
by assuming that if the incident ionising flux increased by a factor $x$, the
cloud would need a hydrogen column density greater by the same factor $x$ to
remaining optically thick. \citet{per03} showed that the cumulative number of
QSO absorbers with hydrogen column densities greater than $N_H$ goes as
$N_H^{-0.3}$. Using these figures, the number of externally ionised clouds
should decrease by a factor of $\sim 2$: not enough to explain our 
non-detection. There is no easy way to estimate the size of this effect on
internally ionised clouds.

The extremely strong UV emission from PKS 0424-131 could physically destroy 
dwarf galaxies in its vicinity, as described by \citet{bar99}, \citet{ben02} 
and references therein. This photoevaporation mechanism heats the gas in
small dark matter halos so much that it becomes unbound. This destruction may
have taken place at a much higher redshift, during the assembly of the very 
massive black hole we infer in this quasar. This mechanism has the advantage of
explaining the absence of Ly$\alpha$ emitters around our UV bright source but
not around radio galaxies, and could well destroy internally ionised Ly$\alpha$
sources, if they are dwarf galaxies residing in small dark matter halos.

\section{Conclusions}

Why then did we see nothing? There are several possible reasons which might
explain the lack of externally ionised clouds, but the lack of internally
ionised clouds is harder to explain. The difference between this UV-bright
quasar and radio galaxies at comparable wavelengths points to the UV 
emission as a possible explanation. We therefore tentatively conclude that the
UV photo-evaporation model seems the best explanation for our non-detection.
With observations of only one QSO, however, this must be regarded as tentative
at best.

How could we test this conclusion? Observations of more QSOs, and a
better understanding of the background density of internally ionised
Ly$\alpha$ sources at this redshift will clearly be vital to
establishing whether this deficit of gas near luminous QSOs is real.
With the recent decommissioning of TTF, this will have to be done at a
different telescope.

If the sensitivity can be increased by a factor of $\sim 3$, we should
be able to detect neutral hydrogen clouds ionised only by the UV
background: ie. we can dispense with a nearby QSO. This should be quite
feasible with long exposure times on an 8-10m class telescope. Alas
most such telescopes do not have imagers capable of achieving the
necessary spectral resolution, and most detectors have poor quantum
efficiency in the UV. One exception will be the Osiris instrument,
which incorporates many of the features of TTF, currently under
construction for the Gran Telescopio Canarias \citep[][]{cep03}. Not 
only should this allow us to image neutral hydrogen clouds far from QSOs and
hence measure their sizes and shapes, the anticipated
sensitivity of this machine should allow many tens of QSO fields to be
surveyed out to $z\sim 5$ over a range of intrinsic ionizing
luminosities.  This may be a relatively direct way to establish the
critical ionizing flux at which star formation is effectively
suppressed in the local universe.

\section*{Acknowledgments}

We would like to thank Frank Briggs for many interesting conversations, Steve
Lee for identifying a suitable calibration lamp for us, and
all the people whose hard work and generosity allowed us to return to work 
at Mt Stromlo Observatory so soon after the January 2003 bushfires. The referee,
Gerry Williger, kindly located archival HST data for us.
This research has made use of the NASA/IPAC Extragalactic Database (NED) 
which is operated by the Jet Propulsion Laboratory, California Institute 
of Technology, under contract with the National Aeronautics and Space 
Administration. It also makes use of data products from the Two Micron 
All Sky Survey, which is a joint project of the University of Massachusetts 
and the Infrared Processing and Analysis Center/California Institute of 
Technology, funded by the National Aeronautics and Space Administration 
and the National Science Foundation. It also make use of data from the
Multimission Archive at the Space Telescope Science Institute (MAST). STScI is
operated by the Association of Universities for Research in Astronomy, Inc,
under NAS contract NAS5-26555.

\bsp

\label{lastpage}


\begin{thebibliography}{99}

\bibitem[Adelberger et al.(2003)]{ade03} Adelberger, K.L., Steidel, C.C.,
Shapley, A.E. \& Pettini, M. 2003, ApJ, 584, 45

\bibitem[Aragon-Salamanca et al.(1994)]{ara94} Aragon-Salamanca, A., Ellis, R.S., 
Schwartzenberg, J.-M. \& Bergeron, J.A. 1994, ApJ, 421, 27

\bibitem[Barkana \& Loeb(1999)]{bar99} Barkana, R. \& Loeb, A. 1999, ApJ, 523,
54

\bibitem[Barthel et al.(1988)]{bar88} Barthel, P.D., Miley, G.K., Schilizzi,
R.T. \& Lonsdale, C.J. 1988, A\&A Suppl., 73, 515

\bibitem[Benson et al.(2002)]{ben02} Benson, A.J., Lacey, C.G., Baugh, C.M.,
Cole, S. \& Frenk, C.S. 2002, MNRAS, 333, 156

\bibitem[Bergeron \& Boiss\'{e}(1991)]{ber91} Bergeron, J. \& Boiss\'{e}, P.
1991, A\&A, 243, 344

\bibitem[Bergeron et al.(1999)]{ber99} Bergeron, J., Petitjean, P., Cristiani,
S., Arnouts, S., Bresolin, R. \& Fasano, G. 1999, A\&A, 343, L40

\bibitem[Bertin \& Arnouts(1996)]{ber96} Bertin, E. \& Arnouts, S. 1996, A\&AS, 117, 393

\bibitem[Bicknell, Saxton \& Sutherland(2003)]{bic03} Bicknell, G.V., Saxton,
C.J. \& Sutherland, R.S. 2003, PASA, 20, 102

\bibitem[Bland-Hawthorn, Sokolowski, \& Cecil(1991)]{bla91} Bland-Hawthorn, J., Sokolowski, J. \& Cecil, G.N. 1991, ApJ, 375, 78

\bibitem[Bland-Hawthorn \& Jones(1998)]{bla98} Bland-Hawthorn, J. \& Jones, D.H. 1998, PASA, 15, 44

\bibitem[Bruscoli et al.(2003)]{bru03} Bruscoli, M., Ferrara, A., Marri, S., Schneider, R., Maselli, A.,
Rollinde, E. \& Aracil, B. 2003, MNRAS 343, 41

\bibitem[Bunker, Marleau \& Graham(1998)]{bun98} Bunker, A.J., Marleau, F.R. \&
Graham, J.R., 1998, AJ, 116, 2086

\bibitem[Campos et al.(1999)]{cam99} Campos, A., Yahil, A., Windhorst, R.A.,
Richards, E.A., Pascarelle, S., Impey, C. \& Petry, C. 1999, ApJL, 511, L1

\bibitem[Cayatte et al.(1990)]{cay90} Cayatte, V., Balkowski, C., van Gorkom, J. H., \& 
Kotanyi, C. 1990, AJ, 100, 604

\bibitem[Cepa et al.(2003)]{cep03} Cepa, J. et al. 2003, SPIE, 4841, 1739

\bibitem[Chen et al.(2001)]{che01}
Chen, H.-W., Lanzetta, K.M., Webb, J.K. \& Barcons, X.2001, ApJ, 559, 654

\bibitem[Condon et al.(1998)]{con98} Condon, J.J., Cotton, W.D., Greisen, E.W.,
Yin, Q.F., Perley, R.A., Taylor, G.B., \& Broderick, J.J. 1998, AJ, 115, 1693

\bibitem[Cowie \& Hu(1998)]{cow98} Cowie, L.L. \& Hu, E.M. 1998, AJ, 115, 1319

\bibitem[Croft et al.(2002)]{cro02} Croft, R.A.C., Hernquist, L., Springel, V., Westover, M.
\& WHite, M.  2002, ApJ, 580, 634

\bibitem[Crotts \& Fang(1998)]{cro98} Crotts, A.P.S. \& Fang, Y. 1998, ApJ, 502, 16

\bibitem[Espey et al.(1989)]{esp89} Espey, B.R., Carswell, R.F., Bailey, J.A., 
Smith, M.G. \& Ward, M.J. 1989, ApJ, 342, 666

\bibitem[Fall \& Pei(1993)]{fal93} Fall, S.M. \& Pei, Y.C. 1993, ApJ, 402, 479

\bibitem[Ferrarese(2002)]{fer02} Ferrarese, L. 2002, ApJ, 578, 90

\bibitem[Foltz et al.(1984)]{fol84} Foltz, C.B., Weymann, R.J., Roser, H.-J. \& 
Chaffee, F.H.Jr. 1984, ApJL, 281, 1

\bibitem[Francis, Wilson \& Woodgate(2001)]{fra01} Francis, P.J., Wilson, G.M.
\& Woodgate, B.E. 2001, PASA, 18, 64

\bibitem[Francis \& Hewett(1993)]{fra93} Francis, P.J. \& Hewitt, P.C. 1993, AJ,
105, 1633

\bibitem[Francis \& Williger(2004)]{fra04} Francis, P.J. \& Williger, G.M. 2004,
ApJL, 602, L77

\bibitem[Francis et al.(1991)]{fra91} Francis, P.J., Hewett, P.C., Foltz, C.B.,
Chaffee, F.H., Weymann, R.J. \& Morris, S.L. 1991, ApJ, 373, 465

\bibitem[Francis et al.(2001)]{fra01b} Francis, P.J. et al. 2001, ApJ, 554, 1001

\bibitem[Giovanelli \& Haynes(1985)]{gio85} Giovanelli, R. \& Haynes, M.P. 1985, ApJ, 292, 404 

\bibitem[Gould \& Weinberg(1996)]{gou96} Gould, A. \& Weinberg, D.H. 1996,
ApJ, 468, 462

\bibitem[Hogan \& Weymann(1987)]{hog87} Hogan, C.J. \& Weymann, R.J. 1987,
MNRAS, 225, P1

\bibitem[Haehnelt, Steinmetz \& Rauch(1998)]{hae98} Haehnelt, M.G., Steinmetz,
M.G. \& Rauch, M. 1998, ApJ, 495, 647

\bibitem[Haiman \& Rees(2001)]{hai01} Haiman, Z. \& Rees, M.J. 2001, ApJ, 468,
87

\bibitem[Hu \& McMahon(1996)]{hu96} Hu, E.M. \& McMahon, R.G. 1996, Nature, 382,
231

\bibitem[Hu, McMahon \& Egami(1996)]{hu96a} Hu, E.M., McMahon, R.G. \& Egami, E.
1996, ApJL, 459, 53

\bibitem[Hu et al.(2004)]{hu04} Hu, E.M., Cowie, L.L., Capak, P., McMahon, R.G.,
Hayashine, T. \&  Komiyami, Y. 2004, AJ, 127, 563

\bibitem[Impey et al.(2002)]{imp02} Impey, C.D., Petry, C.E., Foltz, C.B.,
Hewett, P.C. \& Chaffee, F.H. 2002, ApJ, 574, 623

\bibitem[Jakobsen et al.(2003)]{jak03} Jakobsen, P., Jansen, R.A., Wagner, S. \& Reimers, D.
2003, A\&A, 397, 891

\bibitem[Jones, Shopbell \& Bland-Hawthorn(2002)]{jon02} Jones, D.H., Shopbell,
P.L. \& Bland-Hawthorn, J. 2002, MNRAS, 329, 759

\bibitem[Kaspi et al.(2000)]{kas00} Kaspi, S., Smith, P.S., Netzer, H., Maoz,
D., Jannuzi, B.T. \& Giveon, U. 2000, ApJ, 533, 631

\bibitem[Kollmeier et al.(2003)]{kol03} Kollmeier, J.A., Weinberg, D.H., Dav\'{e}, R.
\& Katz, N. 2003, ApJ, 594, 75

\bibitem[Kurk et al.(2000)]{kur00} Kurk, J.D., R\"{o}ttgering, H.J.A., 
Pentericci, L., Miley, G.K., van Breugel, W., Carilli, C.L., Ford, H., Heckman,
T., McCarthy, P. \& Moorwood, A. 200, A\&A, 358, L1

\bibitem[Kudritzki et al.(2000)]{kud00} Kudritzki, R.-P., M\'{e}ndez, R.H.,
Feldmeier, J.J., Ciardullo, R., Jackoby, G.H., Freeman, K.C., Arnaboldi, M.,
Capaccioli, M., Gerhard, O. \& Ford, H.C. 2000, ApJ, 536, 19

\bibitem[Le Fevre et al.(1996)]{lef96} Le Fevre, O., Deltorn, J.M., Crampton, D.
\& Dickinson, M. 1996, ApJL, 471, L11

\bibitem[Liske \& Williger(2001)]{lis01} Liske, J. \& Williger, G.M. 2001, MNRAS, 328, 353

\bibitem[Maloney \& Bland-Hawthorn(2001)]{mal01} Maloney, P.R. \& Bland-Hawthorn, J. 2001, ApJ, 553, L129

\bibitem[Martini \& Schneider(2003)]{mar03} Martini, P. \& Schneider, D.P. 2003, ApJL, 597, L109

\bibitem[McIntosh et al.(1999)]{mci99} McIntosh, D.H., Rieke, M.J., Rix, H.-W.,
Foltz, C.B. \& Weymann, R.J. 1999, ApJ, 514, 40

\bibitem[Miller, Goodrich \& Mathews(1991)]{mil91} Miller, J.S., Goodrich, R.W. \& Mathews, W.G. 1991, ApJ, 378, 47

\bibitem[M\o ller \& Warren(1993)]{mol93} M\o ller, P. \& Warren, S.J. 1993, 
A\&A, 270, 43

\bibitem[M\o ller et al.(2002)]{mol02} M\o ller, P., Warren, S.J.,
Fall, S.M., Fynbo, J.U. \& Jakobsen, P. 2002, ApJ, 574, 51

\bibitem[Ouchi et al.(2003)]{ouc03} Ouchi, M. et al. 2003, ApJ, 582, 60

\bibitem[Palunas et al.(2004)]{pal04} Palunas, P., Teplitz, H.I., Francis, P.J.,
Williger, G.M. \& Woodgate, B.E. 2004, ApJ, 602, 545

\bibitem[Pascarelle et al.(1996)]{pas96} Pascarelle, S.M., Windhorst, R.A.,
Driver, S.P., Ostrander, E.J. \& Keel, W.C. 1996, ApJL, 456, L21

\bibitem[Pascarelle et al.(1998)]{pas98} Pascarelle, S.M., Windhorst, R.A. \&
Keel, W.C. 1998, AJ, 116, 2656

\bibitem[Pentericci et al.(2000)]{pen00}  
Pentericci, L., Kurk, J.D., R\"{o}ttgering, H.J.A.,
Miley, G.K., van Breugel, W., Carilli, C.L., Ford, H., Heckman,
T., McCarthy, P. \& Moorwood, A. 2000, A\&A, 361, 25

\bibitem[Pei, Fall \& Hauser(1999)]{pei99} Pei, Y.C., Fall, S.M. \& Hauser, M.G.
1999, ApJ, 522, 604

\bibitem[P\'{e}roux et al.(2003)]{per03} P\'{e}roux, C., McMahon, R.G.,
Storrie-Lombardi, L.J. \& Irwin, M.J. 2003, MNRAS, 346, 1103

\bibitem[Petitjean, Rauch \& Carswell(1994)]{pet94} Petitjean, P., Rauch, M.
\& Carswell, R.F. 1994, A\&A, 291, 29

\bibitem[Prochaska \& Wolfe(1998)]{pro98} Prochaska, J. \& Wolfe, A. 1998, ApJ,
507, 113

\bibitem[Rauch, Sargent \& Barlow(1999)]{rau99} Rauch, M., Sargent, W.L.W. \& 
Barlow, T.A. 1999, ApJ, 515, 500

\bibitem[Rosati, Borgani \& Norman(2002)]{ros02} Rosati, P., Borgani, S. \&
Norman, C. 2002, ARAA, 40, 539

\bibitem[Scannapieco \& Oh(2004)]{sca04} Scannapieco, E. \& Oh, S.P. 2004, ApJ
submitted (astro-ph/0401087)

\bibitem[Schlegel, Finkbeiner \& Davis(1998)]{sch98} Schlegel, D.J.,
Finkbeiner, D.P. \& Davis, M. ApJ, 500, 525

\bibitem[Scott et al.(2002)]{sco02} Scott, J., Bechtold, J., Morita, M.,
Dobrzycki, A. \& Kulkarni, V.P. 2002, ApJ, 571, 665

\bibitem[Smette et al.(1995)]{sme95} Smette, A., Robertson, J.G., Shaver, P.A., Reimers, D., 
Wisotzki, L. \& Koehler, T. 1995, A\&A 113, 199

\bibitem[Solanes et al.(2001)]{sol01} Solanes, J. M., Manrique, A., García-Gómez, C., González-Casado, G., Giovanelli, R., \& Haynes, M. P. 2001, ApJ, 548, 97

\bibitem[Steffen et al.(2003)]{ste03} Steffen, A.T., Barger, A.J., Cowie, L.L.,
Mushotzky, R.F. \& Yang, Y. 2003, ApJL, 596, L23

\bibitem[Steidel et al.(2000)]{ste00} Steidel, C.C., Adelberger, K.L., Shapley,
A.E., Pettini, M., Dickinson, M. \& Giavalisco, M. 2000, ApJ, 532, 170

\bibitem[Steidel et al.(2002)]{ste02} Steidel, C.C., Kollmeier, J.A., Shapley,
A.E., Churchill, C.W., Dickinson, M. \& Pettini, M. 2002, ApJ, 570, 526

\bibitem[Venemans et al.(2002)]{ven02}  Venemans, B.P., 
Kurk, J.D., Miley, G.K., R\"{o}ttgering, H.J.A.,
van Breugel, W., Carilli, C.L., Ford, H., Heckman,
T., McCarthy, \& Pentericci, L.,  2002, A\&A, 361, 25

\end{thebibliography}
\end{document}